\documentclass[aip, jap, 10pt, groupedaddress, twocolumn]{revtex4-1}
\usepackage[T1]{fontenc}
\usepackage{bm}
\usepackage{graphicx}
\usepackage{amsmath}
\usepackage{amssymb}
\usepackage{sidecap}
\usepackage{dcolumn}
\usepackage{color}
\usepackage{epstopdf}
\usepackage{ifpdf}
\usepackage{sidecap}

\begin{document}
\title{\textbf{Veselago lensing in graphene with a p-n junction: classical versus quantum effects}}
\author{S. P. Milovanovi\'{c}}\email{slavisa.milovanovic@uantwerpen.be}
%\affiliation{Departement Fysica, Universiteit Antwerpen \\
%Groenenborgerlaan 171, B-2020 Antwerpen, Belgium}

\author{D. Moldovan}\email{dean.moldovan@uantwerpen.be}
%\affiliation{Departement Fysica, Universiteit Antwerpen \\
%Groenenborgerlaan 171, B-2020 Antwerpen, Belgium}

\author{F.~M.~Peeters}\email{francois.peeters@uantwerpen.be}
\affiliation{Departement Fysica, Universiteit Antwerpen \\
Groenenborgerlaan 171, B-2020 Antwerpen, Belgium}

\begin{abstract}
The feasibility of Veselago lensing in graphene with a p-n junction is investigated numerically for realistic injection leads.
Two different set-ups with two narrow leads are considered with absorbing or reflecting side edges.
This allows us to separately determine the influence of scattering on electron focusing for the edges and the p-n interface.
Both semiclassical and tight-binding simulations show a distinctive peak in the transmission probability that is attributed to the Veselago lensing effect.
We investigate the robustness of this peak on the width of the injector, the position of the p-n interface and different gate potential profiles.
Furthermore, the influence of scattering by both short- and long-range impurities is considered.
\end{abstract}


\date{Antwerp, \today}

\maketitle

\section{Introduction}
\label{sint}

Graphene, the first two-dimensional atomic thin material ever made\cite{c1}, exhibits several exotic properties.
It has exceptional mechanical properties with a Young modulus of the order of tera-Pascal\cite{c2}.
Furthermore, a very large thermal conductivity of suspended graphene was reported\citep{c3}.
Even though it is just one-atom thick, graphene absorbs $2.3 \%$ of white light\cite{c4} which makes it suitable for photodetectors.
Moreover, due to its metallic properties graphene has been proven to be a good candidate for transparent electrodes.

However, most prominent are its electronic properties.
Graphene is a zero-gap semiconductor in which conduction and valence bands touch in six points of the Brillouin zone resulting in a low-energy linear dispersion and hence its particles behave like ultra-relativistic Dirac fermions.
This is the origin for many extraordinary properties including Klein tunneling\cite{c4a}, the anomalous Hall effect observed at room temperature\cite{c5}, minimum conductivity\cite{c5, c6}, and high carrier mobility\cite{c7}, to name a few.

\begin{figure*}[htbp]
	\includegraphics[width=17cm]{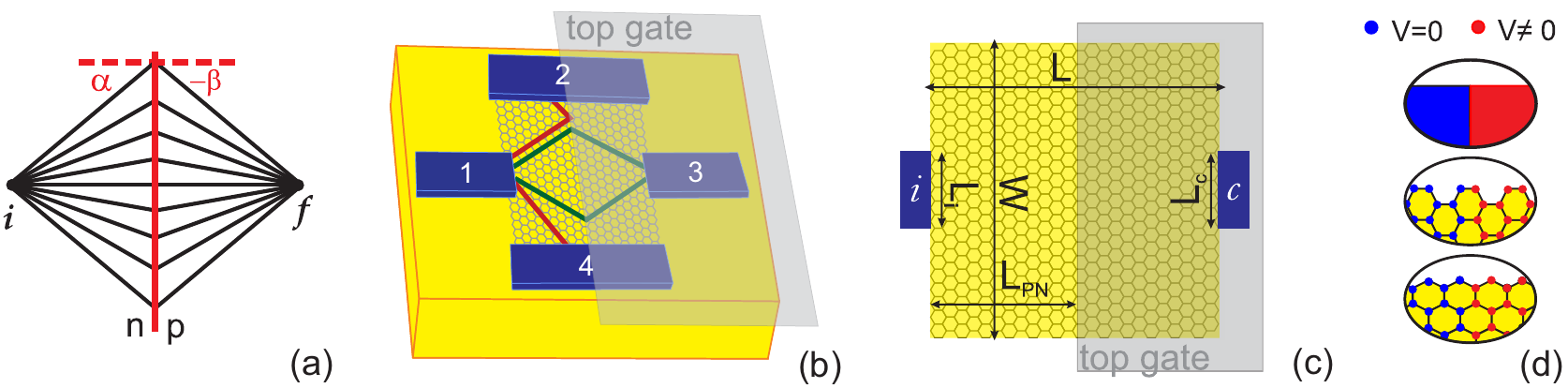}
	\caption{(Color online) (a) Perfect focusing of the electron beam injected from a point injector $i$ into a focal point $f$ due to the negative refraction at a p-n junction in graphene. (b) Four-terminal system.
	Red and green lines are classical trajectories of electrons injected from a narrow lead $1$, scattered on a p-n interface positioned at $L_{PN}$ from the injector and collected at lead 3 (green lines).
	Wide terminals 2 and 4 collect all the electrons that are not focused towards lead 3 (red lines). (c) Two-terminal system.
	The output signal is affected by scattering at the edges. (d) Different types of edges (from top to bottom): continuous (classical calculations), armchair, and zigzag.} 
	\label{f1}
\end{figure*}

It has been shown that by applying a pair of gates one can form a p-n junction in graphene\cite{c8}.
These bipolar devices open the door for novel graphene-based electronics.
High carrier mobility is responsible for a large mean free path (of the order of a few micrometers\cite{c8a}) and allows the manufacturing of micron-size ballistic devices.
Graphene p-n junctions have been used in various experiments to demonstrate many classical phenomena including transverse magnetic focusing \cite{c9} and electrical transport with snake states \cite{c10, c11}, as well as quantum mechanical effects such as Klein tunneling\cite{c4a}, integer and fractional quantum Hall effect\cite{c8}, and Aharonov-Bohm oscillations \cite{c13}.

Furthermore, due to its meta-material properties the p-n interface in graphene acts as a focusing lens for electrons\cite{r1}.
In the case of a bipolar junction the negative refraction in graphene will cause transmission of electrons through the p-n interface with a negative angle which can be used to focus an electron beam injected from a point source to a focal point on the opposite side of the junction, as shown schematically in Fig. \ref{f1}(a).
This is called a Veselago lens.
However, a point source is a theoretical concept that is impossible to realize in a real device where the current source always has a finite width.
On the other hand, graphene, due to its aforementioned large mean free path, allows carriers to propagate over large distances while preserving their phase information.
This coherent transport was observed in high-quality graphene encapsulated in h-BN where the interference of Hall edge channels occurred in a micron-sized device\cite{c14}.
Hence, graphene encapsulated in insulating h-BN, due to its extremely clean surface, qualifies as a promising candidate for the experimental observation of the Veselago lensing effect.
Very recently, Lee \textit{et al.} succeeded in observing a signature of negative refraction in graphene \cite{c15}.

The aim of this paper is to investigate the influence of: \textit{i}) the injector's width, \textit{ii}) the presence of scatterers, \textit{iii}) quantum effects, and \textit{iv}) the type of sample edge on Veselago lensing.
We simulate the electron transport in two different systems shown in Figs. \ref{f1}(b) and (c) that can be realized experimentally.
In both systems electrons are injected from a finite width injector \textit{i} (in Fig. \ref{f1}(b) the injector is lead 1) and are collected at the opposite side of the p-n junction, at collector \textit{c} (in Fig. \ref{f1}(b) lead 3 acts as the collector).
Additionally, one of the systems has two wide leads (leads 2 and 4 in Fig. \ref{f1}(b)) added perpendicular to the p-n interface while the other system (Fig. \ref{f1}(c)) has reflecting edges.
This will allow us to disentangle effects due to scattering on the p-n junction and those resulting from boundary scattering.
We will present results based on both semiclassical and quantum mechanical approaches which allows us to assess the importance of quantum effects.
The semiclassical approach is based on the well-known billiard model\cite{r110} where the transmission through the p-n interface is obtained from quantum mechanics.
Further information about the implementation of this model can be found in Ref. \onlinecite{r3}.
Tight-binding calculations are performed using the KWANT package\cite{ckwant} where electron transport is treated quantum mechanically.

The paper is organized as follows.
In Sect. \ref{s4ter} we calculate the transmission probabilities in the four-terminal system illustrated in Fig. \ref{f1}(b).
We examine the influence of the Fermi energy, width of the injector, position of the p-n interface and gate potential on electric transport using the methods mentioned above.
In Sec. \ref{s2ter} the same calculations are then performed for the system in Fig. \ref{f1}(c).
In Sect. \ref{simp} we investigate the influence of impurities on the Veselago lensing effect.
Our final remarks and conclusions are given in Sect. \ref{sconc}.

\section{Four-terminal device}
\label{s4ter}

A schematic of the device is shown in Fig. \ref{f1}(b).
The device is a $W\times L$ graphene flake with a top gate that controls the density of the region below it.
Four leads are added to the system in such a way that two narrow leads are placed parallel to the p-n interface and two wide leads stretch along the length of the system.
To study Veselago focusing we inject electrons from a narrow injector at the non-gated side of the p-n junction (lead 1 in Fig. \ref{f1}(b)) while lead 3 is used as a collector.
Leads 2 and 4 are added to collect electrons that do not reach the p-n interface or that are scattered by the edges of the device or by the p-n interface itself (e.g. the red trajectories in Fig. \ref{f1}(b)).
We are interested to learn how the transmission coefficient $T$, i.e. the probability for an electron injected from lead 1 to be collected at lead 3, depends on different parameters of the system.
Furthermore, due to the symmetry of the system the transmission probabilities for side-leads will be equal and we have $T_2 = T_4 = T_{SL}$. 

\begin{figure*}[htbp]
	\includegraphics[width=17cm]{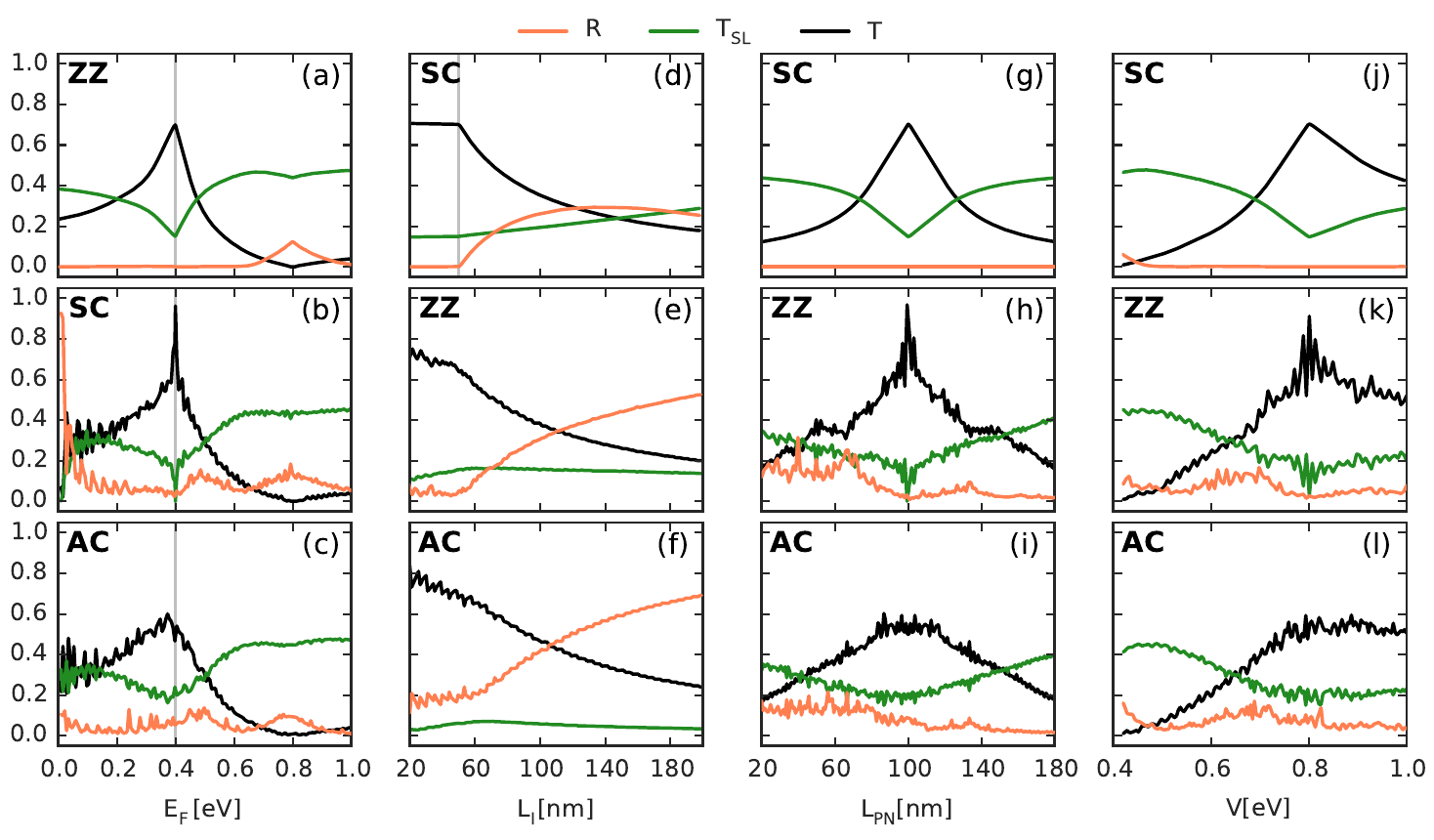}
	\caption{(Color online) Transmission probabilities versus $E_F$, $L_I$, $L_{PN}$, and $V$ for the graphene p-n junction shown in Fig. \ref{f1}(b).
	We compare results from semiclassical (SC) simulations (top row) with the quantum-mechanical tight-binding simulations for zigzag (ZZ) (central row) and armchair (AC) (bottom row) edges at the p-n interface.
	The sum of all transmission probabilities is normalized to unity.}
	\label{f2}
\end{figure*}

The results of our calculations are presented in Fig. \ref{f2}.
In the top row we show the results of the semiclassical (SC) calculation.
The next two rows shows the quantum mechanical results obtained using the tight-binding method where the injector and collector (leads 1 and 3) have armchair (AC) and zigzag (ZZ) edges, presented in the central and bottom rows, respectively.
We should mention that in the case of the tight-binding calculations adding leads along the length of the flake makes the type of side edges irrelevant.
However, the type of edge at the pn-interface turns out to be important.
In the case of armchair edges the pn-interface will be zigzag and vice versa, as shown in Fig. \ref{f1}(d).
This will have a significant impact on the transmission through the p-n interface and the two cases show different responses as shown in Fig. \ref{f2}.
The results shown in these figures were obtained with the following parameters: $W = L = 200$ nm, $L_c = 50$ nm, $L_I = 50$ nm (except for Figs. \ref{f2}(d-f)), $E_F = 0.4$ eV (except for Figs. \ref{f2}(a-c)), $V = 0.8$ eV (except for Figs. \ref{f2}(e, j-l)), and $L_{PN} = L/2$ (except for Figs. \ref{f2}(g-i)) where the definition of $W, L, L_I, L_c, L_{PN}$ are shown in Fig. \ref{f1}(c), while $E_F$ and $V$ are the Fermi energy and the top gate potential, respectively.

\begin{figure*}[htbp]
	\includegraphics[width=15.5cm]{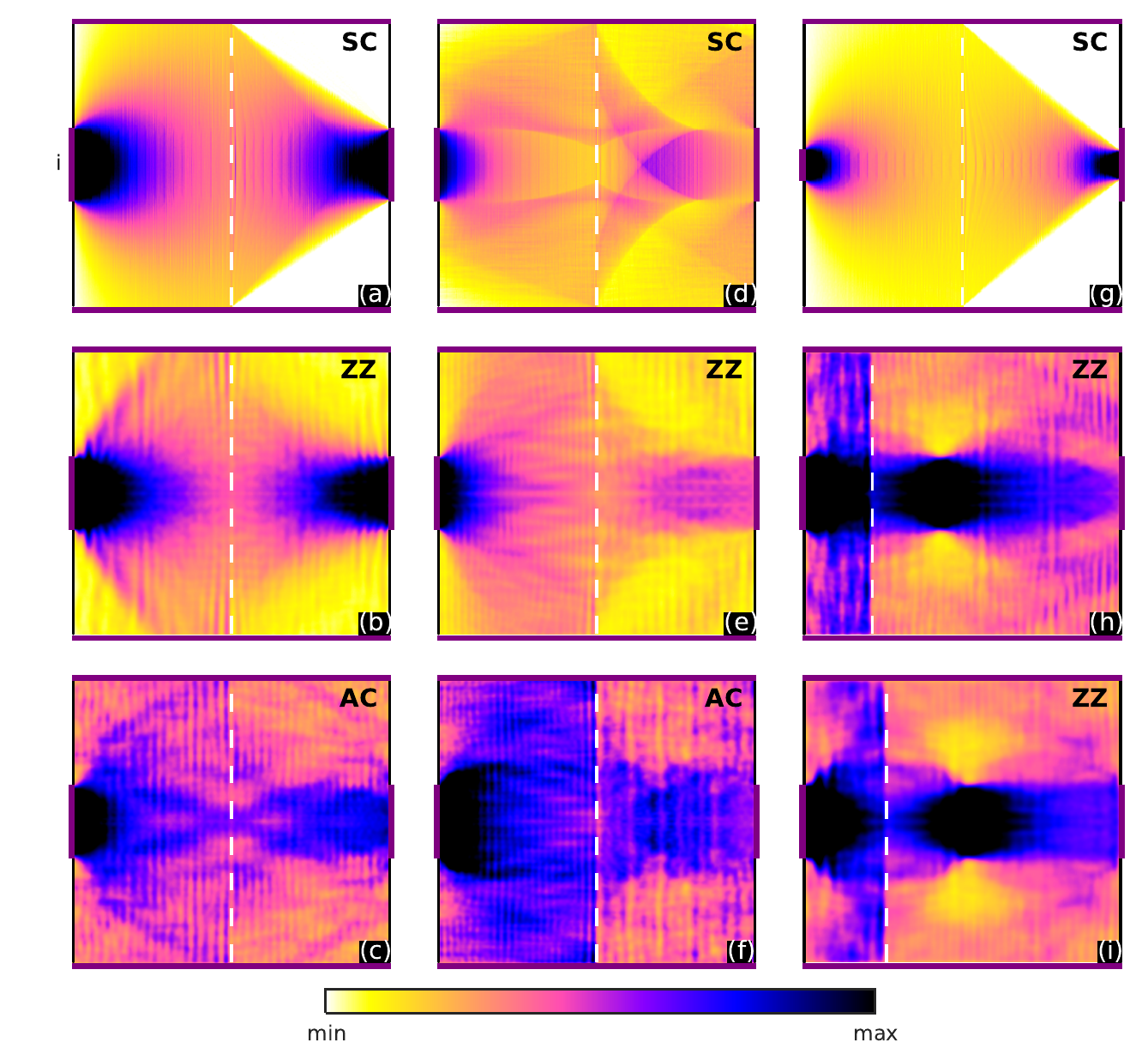}
	\caption{(Color online) Current density plots for the system shown in Fig. 1(b). Plots are made using the following parameters: (a - c) $E_F = 0.4$ eV and $V = 0.8$ eV (except (b) where we used $E_F = 0.39$ eV), (d-f) $E_F = 0.48$ eV, (g) $L_I = 20$ nm, (h) $L_{PN} = 43$ nm, and (i) $L_{PN} = 52$ nm. All other parameters are given in Sect. \ref{s4ter}. Labels SC, AC, and ZZ refer to semiclassical system, system with armchair, and zigzag p-n interface, respectively.}
	\label{f3}
\end{figure*}

In Figs. \ref{f2}(a-c) we show the transmission probabilities versus the Fermi energy as obtained from our semiclassical calculations and armchair/zigzag flakes obtained using the tight-binding approach, respectively.
All figures show a distinctive peak at the same value of the Fermi energy, $E_F = V/2 = 0.4$ eV (shown in figures by the gray vertical line).
This peak is a consequence of the Veselago lensing in graphene.
Namely, electrons that transmit through the pn-interface will be focused in a focal point on the opposite side of the pn-junction as shown in Fig. \ref{f1}(a).
However, due to the finite width of the injector instead of a focal point we rather have a focal spot.
This is seen in Figs. \ref{f3}(a-c) where we plot the current density for $E_F = 0.4$ eV (except Fig. \ref{f3}(b) where $E_F = 0.38$ eV is used) and $V = 0.8$ eV for the semiclassical, calculation with zigzag, and armchair p-n interface, respectively.
For this value of applied potential and $L_{PN} = L/2$ we have a symmetrical p-n junction where the focal spot is placed at the same distance from the p-n interface as the injector which in this case is at the position of the collector.
Fig. \ref{f3}(a) shows our semiclassical calculation for this set of parameters.
We see that all electrons that transmit to the p-side of the junction are focused into the collector.
Furthermore, reflection in this case is zero because of Snell's law, 
\begin{equation}\label{esl}
E_F\sin\alpha=\left(E_F-V\right)\sin\beta, 
\end{equation}
which dictates that in this case we have $\alpha=-\beta$, where $\alpha$ and $\beta$ are the angles of incidence and transmission, respectively, as shown in Fig. \ref{f1}(a).
The transmission probability through the p-n interface, $T_{PN}$, is calculated using Eq. (2) from Ref. \onlinecite{r3} which for $\alpha=-\beta$ results in $T_{PN} = 1$.
Hence, all injected electrons will be either transmitted to the p-side of the junction or collected by the side-leads.
Transmission probabilities, shown in Fig. \ref{f2}(a), tell us that 70$\%$ of injected electrons reach the collector while about $15\%$ is collected by each side-lead.
The reflection is close to zero for a wide range of energies because of the relatively narrow injector and large side-leads which accumulate most of the electrons that do not reach the collector.
This is seen in Fig. \ref{f3}(d) where we plotted the current density for $E_F = 0.48$.
We see that the focal spot is now moved towards the p-n interface and carriers are spread around the whole p-region, however most of them are accumulated by the side-leads.
The exception is when $E_F$ is around $V$ for which we have strong reflection at the p-n interface (low $T_{PN}$) and hence an increase of the reflection probability. 

In the case of a system with zigzag p-n interface, shown in Fig. \ref{f2}(b), we see some disagreement with the semiclassical results, e.g. the reflection probability differs from zero over the whole range of energies.
This is expected in quantum mechanical calculations due to non-specular reflections at the p-n interface or boundaries of the system which is not present in the semiclassical calculations.
Furthermore, notice the "jumps" in the reflection probability.
The jumps occur whenever the Fermi energy hits the bottom of a new electron subband.
At low energies only a few subbands are occupied and these jumps are more pronounced than for larger energies (e.g. for $E_F = 0.4$ eV there are 21 occupied subbands).
However, an increase of the reflection is seen for a Fermi energy just above the gate potential.
This agreement with the semiclassical results shouldn't surprise us given that the current density plots for the two approaches show almost identical features.
Figs. \ref{f3}(b) and (e) show the current densities for $E_F = 0.39$ eV and $E_F = 0.48$ eV, respectively.
The value of the Fermi energy is slightly lower than half the gate potential which is chosen because for $L_{PN} = L/2$ and $E_F = V/2$ a small gap opens in the spectrum of the leads that are partially covered by the top gate.
This is the reason why in this case the transmission shows the highest peak among Figs. \ref{f2}(a-c).
Approximately $96 \%$ of the injected electrons will be collected by the collector (lead 3) while the rest is reflected back to the injector.
Moreover, due to the low number of available states in the side-leads for energies around $V/2$ electrons reflected at the p-n interface are reflected back to the injector.
Hence, the peak in the reflection appears at $E_F = 0.5$ eV.

Transmission probabilities for the system with armchair p-n interface, given in Fig. \ref{f2}(c), show similarities with previous systems, however, the highest transmission is now only $56\%$.
The reason for the lower $T$ can be found in Figs. \ref{f3}(c) and (f).
Here we see that unlike the case with semiclassical approach and the system with zigzag p-n interface this system doesn't show proper focusing abilities.
Although the focal spot can be observed in both figures we see that the current is spread over the whole system.
These current responses are quite different from what is expected using ray optics and classical trajectories.
Of course, the cause for this can be linked with the specific edges of our system.
However, having in mind that by adding wide leads along the length of the nanoflake, i.e. absorbing boundary conditions, the difference comes from the edge structure at the p-n interface.
In the case of armchair edges (of the injector) the p-n interface is zigzag and consists only of atoms of one sublattice while in the case of zigzag edges the p-n interface is armchair and consists of atoms from both sublattices.

\begin{figure*}[htbp]
	\includegraphics[width=15.0cm]{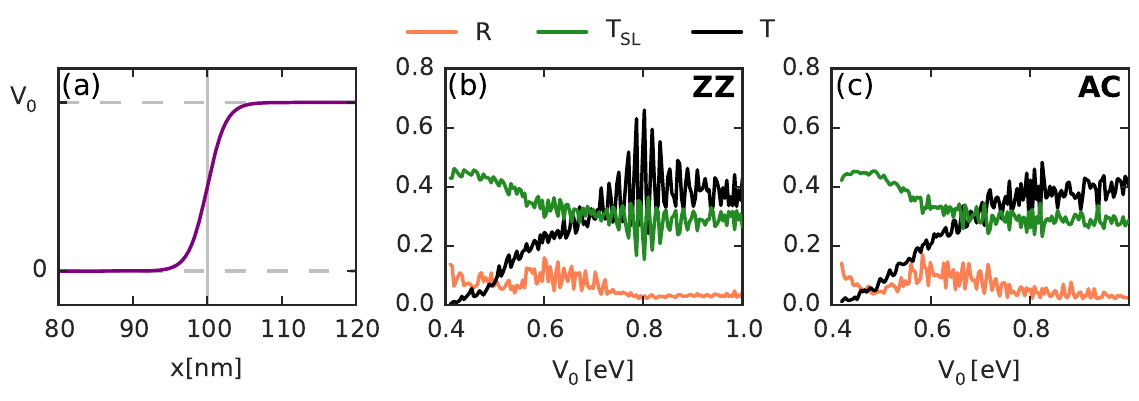}
	\caption{(Color online) (a) Smooth potential profile for the p-n interface. Transmission probabilities for the system with (b) zigzag (ZZ) and (c) armchair (AC) p-n interface versus gate potential.}
	\label{f3a}
\end{figure*}

Next, we study the dependence of the transmission probabilities on the width of the injector.
In these calculations we change the size of the injector while the width of the collector is fixed to $L_c = 50$ nm.
The semiclassical calculation presented in Fig. \ref{f2}(d) shows that all probabilities are constant up to a certain value of $L_I$ (marked with gray vertical line) after which $T$ starts decaying while $R$ and $T_{SL}$ increase.
The gray vertical line is set at $L_I = 50$ nm which is the width of the collector.
This can be understood from the fact that for a p-n interface placed at the middle of the system and $E_F = V/2$ the focal spot is at the position of the collector.
The size of the focal spot is however determined by the width of the injector.
Namely, the p-n interface acts as a lens and the injector is hence mirrored on the opposite side of the junction.
Therefore, for widths of the injector smaller than the width of the collector the same number of injected electrons will reach the collector.
This is shown in Fig. \ref{f3}(g).
Only for the condition $L_I > L_c$ will the focal spot be wider than the collector which causes carriers to scatter on the edges and we see a decrease of the transmission. 

In the case of the zigzag p-n interface shown in Fig. \ref{f2}(e) we used $V = 0.85$ eV, slightly higher than twice the Fermi energy.
This is done because for $E_F = V/2$ a gap opens in the spectrum of the side-leads and $T_{SL}$ is always zero.
The important difference with the semiclassical calculation is that the reflection continuously increases with $L_I$.
The reason is that the number of available states in the side-leads and collector are limited.
As we increase the width of the injector more states appear below the Fermi energy, however since the number of states in the collector is constant additional carriers are either reflected back to the injector or transmitted to the side-leads.
Hence, reflection increases. 

When the system with armchair p-n interface is used the situation is similar.
This is shown in Fig. \ref{f2}(f).
The difference is that now only one band is below the Fermi energy in the side-leads.
This is the reason for the low $T_{SL}$.
Furthermore, for large $L_I$ the transmission $T_{SL}$ decreases.
The same explanation applies for $T$ in the case of a zigzag p-n interface.

Next, we investigate the dependence of the transmission probabilities on the position of the p-n interface $L_{PN}$.
This is shown in Figs. \ref{f2}(g-i) where we shift the p-n interface from the injector to the collector.
In these calculations we again use $V = 2E_F = 0.8$ eV.
Fig. \ref{f2}(a) shows the results of our semiclassical calculations.
First, notice the absence of reflection. $T$ and $T_{SL}$ show the expected behavior.
The position of the focal spot is determined by the value of the gate potential.
Using simple geometry arguments we calculate the position of the focal spot, $\mathit{f}$, as 
\begin{equation}\label{efs}
\mathit{f} = L_{PN}\left| \frac{\tan\alpha}{\tan\beta}\right|,
\end{equation}
where angles $\alpha$ and $\beta$ are shown in Fig. \ref{f1}(a) and obey Eq. \eqref{esl}.
Therefore, when $\mathit{f}$ coincides or is close to the collector's position we will have high transmission while in the other situations reflection at the edges of the system will contribute to an increase of $T_{SL}$. 

\begin{figure*}[htbp]
	\includegraphics[width=17cm]{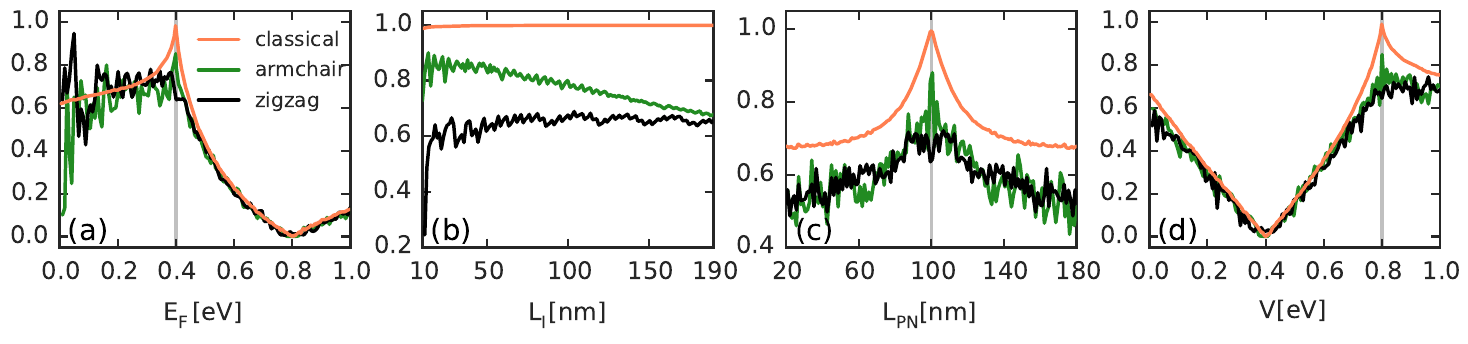}
	\caption{(Color online) Transmission probability versus (a) the Fermi energy, (b) width of the injector, (c) position of the p-n interface, and (d) gate potential for semiclassical, armchair, and zigzag system for the case shown in Fig. \ref{f1}(c).}
	\label{f4}
\end{figure*}

The tight-binding calculations show similar behavior.
However, notice that in the case of the zigzag p-n interface, shown in Fig. \ref{f2}(h), for $L_{PN} = 100$ nm almost all injected electrons are collected by the collector.
Of course, the reason is the band gap in the side-leads.
Other interesting features are the plateaus that appear for $50$ nm $ < L_{PN} < 66$ nm and $133$ $<L_{PN} < 150$ nm.
Notice that the plateaus appear as a consequence of the drop of the reflection while the behavior of $T_{SL}$ doesn't show any peculiarity.
We examine the influence of the Fermi energy, injector's width, width and length of the device on the size and position of the plateaus and conclude that there were no changes except with the length.
If one changes the length of the sample both the size and position of the plateau will change.
However, the plateau always appears in the range of $L/4 < L_{PN} < L/3$ from the injector or collector.
For $L_{PN} < L/3$ the p-n interface is too close to the injector for proper bending of the beam.
Instead, the beam is spread over the n-region as shown in Figs. \ref{f3}(h) and (i).
The figures show the difference between $L_{PN} < L/4$ and $L/4 < L_{PN} < L/3$.
In the first case, shown in Fig. \ref{f3}(h), the focal spot is placed in the same half of the device as the injector.
As a consequence after the focal spot the beam starts to disperse.
While we move the focal spot (i.e. changing of $L_{PN}$) towards $L/2$ transmission is increasing due to the collimation of the beam.
Fig. \ref{f3}(i) depicts the current density for $L/4 < L_{PN} < L/3$ which shows a highly collimated beam after the focal point.
The beam remains collimated in the whole range of position of the p-n interface and hence we have a plateau.
A similar situation occurs when the p-n interface is too close to the collector. 

In the case of the armchair p-n interface, shown in Fig. \ref{f2}(i), we see that the peak around $L_{PN} = L/2$ is much lower than for the case of a zigzag p-n interface and only about $60\%$ of electrons reach the collector.
The reason is not only the absence of a band gap in the side-leads but also the higher reflection back to the injector.
Higher reflection suggests fewer carriers reach the p-region of the junction as compared with the system with zigzag p-n interface, which is also confirmed by the current density plots.
All this suggests that the armchair p-n interface has poorer lensing abilities than the zigzag one. 

Figs. \ref{f2}(j-l) show the transmission probabilities $R$, $T$, and $T_{SL}$ versus the gate potential $V$.
The calculations are performed for $E_F = 0.4$ eV and $L_{PN} = 100$ nm.
The results show the increase of transmission $T$ as the gate potential approaches $V = 0.8$ eV while $T_{SL}$ decreases in this range.
The reflection is low and does not go above $20\%$.
Interestingly, in the case of armchair p-n interface, the transmission doesn't show a peak at $V = 0.8$ eV but rather saturates to $T \approx 55\%$.

Up to now, we have limited ourselves to a step potential for convenience which allowed us to elucidate the physics more clearly.
However, in order to approach a realistic system more closely, we should replace the sharp potential with a smooth transition between the two regions.
We model this by
\begin{equation}
\label{etan}
V(x) = \frac{V_0}{2}\left[1 + \tanh\left(\frac{x - L_{PN}}{\Delta}\right)\right], 
\end{equation}
in the region around the p-n interface.
The potential profile for $\Delta = 2.5$ nm and $V_0 = 0.8$ eV is shown in Fig. \ref{f3a}(a).
In Figs. \ref{f3a}(b) and (c) we present our quantum mechanical results for the transmission probabilities as function of the gate potential for armchair and zigzag systems, respectively.
Comparing these results with the ones obtained using a step potential, presented in Figs. \ref{f2}(k) and (l), we see that all main features are preserved.
Moreover, Fig. \ref{f3a}(b) shows a broadening of the peak around $V = 0.8$ eV and a drop of the transmission probability.
While in Fig. \ref{f2}(k) the transmission maximum is higher than $91 \%$, in the case of a smooth potential it drops only to $66\%$.
This is similar for the zigzag system.
Using the step potential the transmission probability goes up to $58\%$ while in the case of a smooth potential this extremum decreases to $46\%$.
Therefore, in order to have better focusing abilities it is essential to construct a device with a very sharp p-n interface.

\section{Two-terminal device}
\label{s2ter}

In this section we investigate how the transmission probabilities are changed if the side-leads are replaced by reflecting boundaries.
Calculations are performed for the system of Fig. \ref{f1}(c) using the same parameters as in the previous section.
The results for the total transmission probability are shown in Fig. \ref{f4}. 

\begin{figure*}[htbp]
	\includegraphics[width=15.5cm]{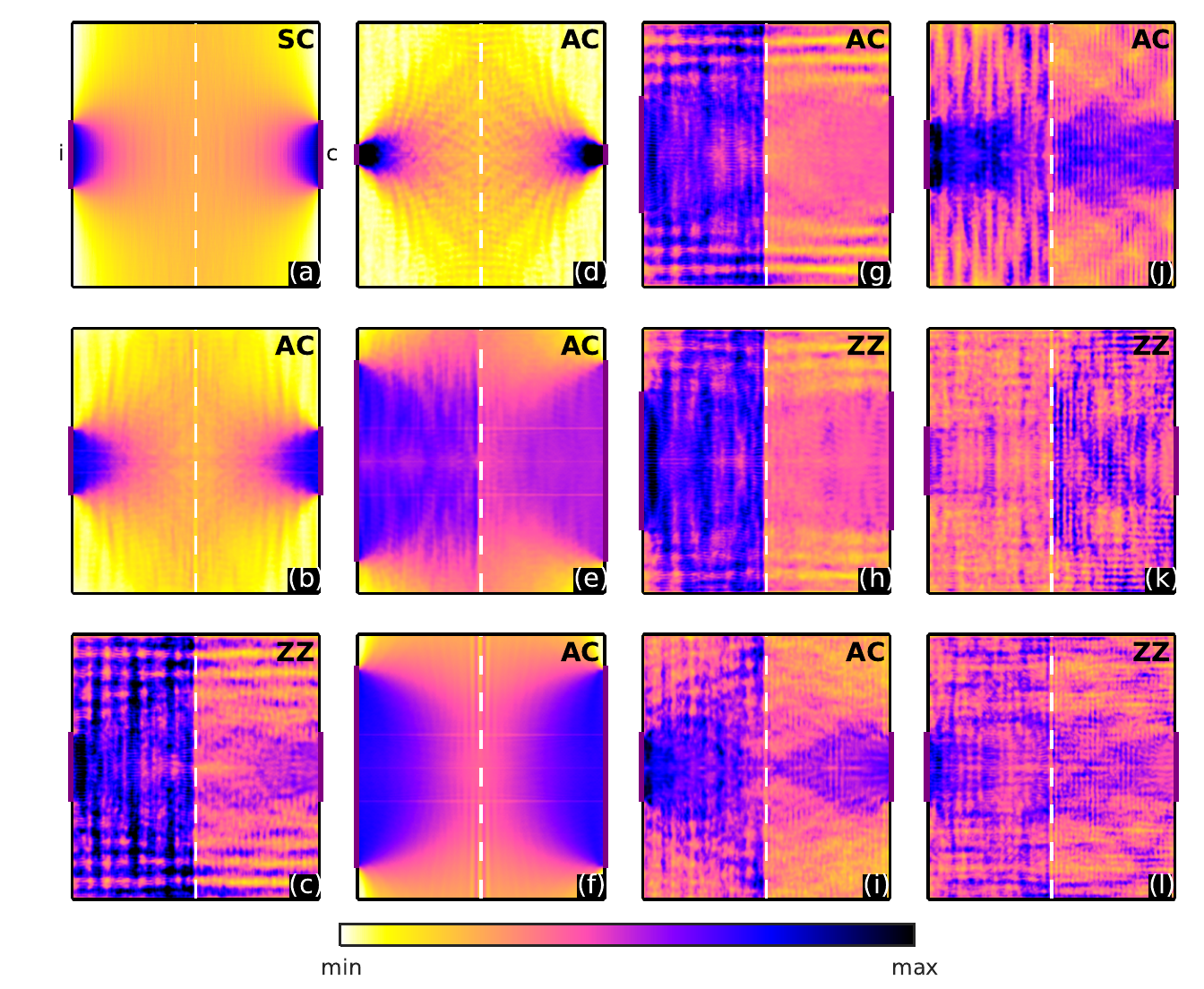}
	\caption{(Color online) Current density plots for the system shown in Fig. 1(c). Plots are made using the following parameters: (a - c) $E_F = 0.4$ eV, (d) $L_I = 13$ nm, (e) $L_I = 150$ nm, (f) $L_I = 150$ nm, (g) $L_I = 86$ nm, (h) $L_I = 103$ nm, (i) $V = 0.9$ eV, (j) $V = 0.95$ eV, (k) $V = 0.9$ eV, and (l) $V = 0.95$ eV. All other parameters are given in Sect. II. Labels SC, AC, and ZZ refer to semiclassical, system with armchair, and zigzag edges at the boundary, respectively.}
	\label{f5}
\end{figure*}

The dependence on the Fermi energy is shown in Fig. \ref{f4}(a).
The orange line in this figure shows the result of our semiclassical calculation.
In the previous section we saw that the reflection was very low (negligible in the case of semiclassical calculation) except for a small range of energies around the gate potential.
This was explained by the fact that the injector is narrow and hence most of the carriers that do not reach the collector are accumulated by the side-leads.
In this case, however, since there are no side-leads, carriers will scatter on the edges until they reach the collector or the injector.
We see that the transmission probability increases as we approach $E_F = V/2$ which means that the focal spot moves towards the collector.
At $E_F = 0.4$ eV the transmission shows a maximum with $99.9\%$ of injected electrons collected by lead $c$.
Similar situation is seen for armchair and zigzag system shown in the same figure by the green and black curves, respectively.
However, since in this range of energies only a small number of bands is occupied, $T$ shows many oscillations.
Nonetheless, in the case of a system with armchair edges transmission exhibits a maximum at $E_F = 0.4$ eV, while in the case of zigzag edges the maximum is achieved at $E_F = 0.05$ eV for which $95\%$ of electrons is transferred to the collector.
We should mention that this energy is just below the bottom of the second energy band (only one energy band is occupied at this point).
In Figs. \ref{f5}(a-c) we plotted the current densities for semiclassical, armchair and zigzag systems, respectively.
We see in Figs. \ref{f5}(a-b) that the focal spot is placed at the position of the collector leading to a maximum in transmission.
For the zigzag system, shown in Fig. \ref{f5}(c), this is not the case.
The focal spot is not clearly seen and furthermore it seems that the current density is higher in the n-region.
This suggests that the cause for the lower transmission is the armchair structure of the p-n interface which diminishes the transmission of the current carriers through it. 

In the range of energies $V/2 < E_F < V$ transmission decreases as a consequence of angle restriction imposed by Snell's law.
From Eq. \eqref{esl} we see that only those angles $\alpha$ for which $\left| E_F/(E_F-V)\right| \sin(\alpha)\leq1$ can transmit through the p-n interface.
As $E_F$ approaches $V$ only a narrow range around $\alpha = 0$ will have a non-zero transmission probability while most electrons are reflected back to the injector.
In the same manner, with increase of energy above this value, the range of allowed angles increases and so does the transmission. 

Next, we study the dependence of the transmission probabilities on the width of the injector and the collector.
In these calculations, unlike in previous section, we simultaneously change the width of injector and collector, i.e. $L_I = L_c$.
The orange curve in Fig. \ref{f4}(b) shows results of our semiclassical calculations.
We see that in this case the transmission probability does not dependent on the width of the injector/collector.
This is not surprising having in mind that in this type of calculation the charge source is perfectly mirrored at the p-side of the junction.
Since the width of the collector is equal to the width of the injector, the injected beam will be transferred to the collector without losses.
However, this is not the case with our tight-binding calculations.
In Fig. \ref{f4}(b) we plot the response of the system with armchair edges by the green curve which shows that the transmission decreases with increase of $L_I$.
When $L_I = L_c = 13$ nm approximately $90\%$ of electrons reach the collector, while for $L_I = L_c = 190$ nm this probability drops to $67\%$.
The reason for this can be understood from the current density plots.
In Figs. \ref{f5}(d) and (e) we show the current densities for $L_I  = L_c = 13$ nm and $L_I = L_c = 150$ nm.
The difference between the two plots is large.
While for the narrow injector we have a perfect focusing pattern with the current source nicely mirrored on the opposite side of the p-n junction, this is not the case for the wide injector.
A possible cause for this lies in the armchair edges of the system.
In the same way that the armchair structure of the p-n interface in the zigzag system diminishes carrier transmission through it, here the armchair edges do not support perfect specular reflections on the walls of the system.
In the system with a narrow injector most electrons are bent on the p-n interface and hence never reach the edges.
For the wide injector, on the other hand, a large number of electrons are injected close to the edges and interact with them and hence distort the focusing pattern.
To confirm this, in Fig. \ref{f5}(f) we plotted the results of the four-terminal system with the same parameters as in Fig. \ref{f5}(e).
Notice that the mirror symmetry in the current profile is restored.
This confirms our assumption that the reflection at the armchair edges is responsible for the absence of the mirror symmetry (with respect to the p-n interface) and hence the lower transmission probability.

\begin{figure}[htbp]
	\includegraphics[width=6.5cm]{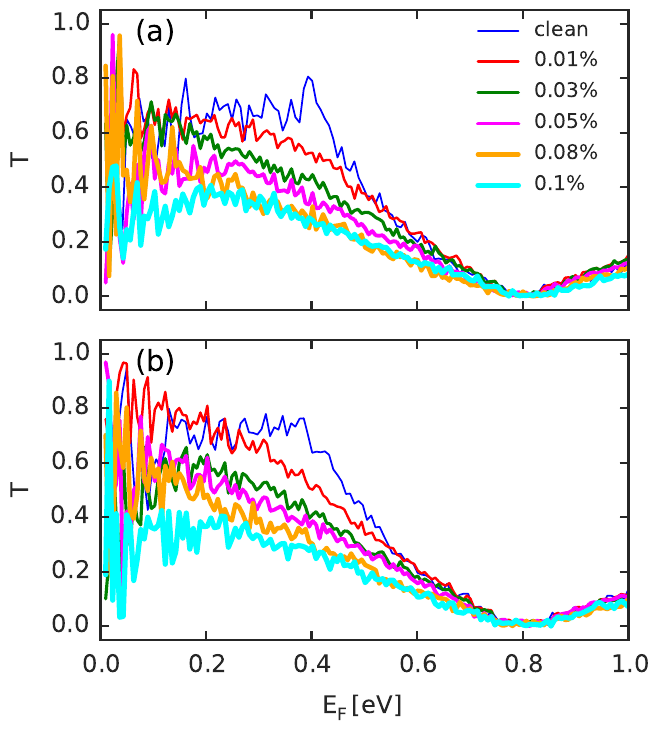}
	\caption{(Color online) Transmission probability versus the Fermi energy for different amount of scatterers indicated in the inset of panel (a). Calculations are performed for (a) armchair and (b) zigzag systems.}
	\label{f7}
\end{figure}

The results of the system with zigzag edges are presented by the black curve in Fig. \ref{f4}(b).
Although not constant the zigzag system shows a slow and periodic change of $T$ (and $R$) with increase of injector's width.
However, we see that roughly $66\%$ of electrons reach the collector, which is almost the same as for the armchair system with wide injector.
Due to the small number of occupied bands oscillations in the transmission appear.
We can use this fact to examine the periodicity of $T$.
Although only 4 minimums are shown in this figure we see that seven new bands appear in-between two minima.
In Figs. \ref{f5}(g) and (h) we plotted the current densities for the first minimum at $L_I = 86.3$ nm and the fist maximum at $L_I = 103$ nm, respectively. 

Fig. \ref{f4}(c) shows the change of transmission probabilities with position of the p-n interface, $L_{PN}$.
This figure shows similar behavior as the ones obtained for the four-terminal systems.
However, notice the difference in the height of the focusing peaks at $L_{PN}  = L/2$.
In the case of our semiclassical calculations, presented in Fig. \ref{f4}(c) by the orange curve, this peak is higher as compared to the four-terminal device.
In Fig. \ref{f2}(g) reflection is zero which means that for $L_{PN}  = L/2$ all electrons that were accumulated by the side-leads in this case end up in the collector.
Of course this is not surprising having in mind the symmetry of the system and the specular reflections on the walls of the system.
On the other hand, in the case of the armchair system this peak is lower as compared to the four-terminal case.
This is shown in Fig. \ref{f4}(c) by the green curve.
As we already mentioned for $L_{PN} = L/2$ a gap opens in the spectrum of the side-leads hence we basically have a two-terminal device.
However, while the peak in Fig. \ref{f2}(h) shows $T > 99\%$, in this case we have "only" $T \approx 90\%$.
This confirms that the reflection on the armchair edge diminishes the Veselago lensing effect in graphene.
The response of a two-terminal system with zigzag edges is shown in Fig. \ref{f4}(c) by the black curve.
We see that in this configuration, as for the semiclassical system, the probabilities are shifted up and furthermore $T > R$ for all values of $L_{PN}$.

Finally, in Fig. \ref{f4}(d) we show the dependence of the transmission probabilities on gate potential for the three systems.
Our semiclassical results exhibit the expected behavior which can be explained similarly as the Fermi energy dependence.
Namely, by changing the gate potential we move the focal spot along the length of the system.
When its position coincides with the position of the collector we have a maximum in the transmission and as we move away from this position the transmission decreases.
However, the tight-binding calculations for the armchair system, for $V> 2E_F$ shows a much slower decrease of the transmission with the gate potential as compared to the semiclassical case.
To understand this irregularity we plotted the current density for $V = 0.9$ eV in Fig. \ref{f5}(i).
With this potential we place the focal spot beyond the collector and hence we should expect reflected current on the wall of the system around the collector since the transmitted beam is wider than the collector (similar for the case when the injector is wider than the collector shown in Fig. \ref{f3}(h)).
Nonetheless, the current density plot doesn't show this behavior.
If we continue to increase the potential to $V = 0.95$ eV, for which there is a small drop of the transmission, the current reflecting on the wall around the collector is again observable, as shown in Fig. \ref{f5}(j).
Therefore, the absence of reflected current is responsible for a slow decrease of the transmission in this region.
In the case of the zigzag system we rather have a saturation of $T$ than a decrease.
This is similar with the situation we had for the four-terminal zigzag device.
In Figs. \ref{f5}(k) and (l) current densities are plotted for $V = 0.9$ eV and $V = 0.95$ eV, respectively.
Notice that the beam focusing pattern is completely destroyed for $V > 2E_F$ and consequently the logic we used to explain the decrease of transmission due to the movement of the focal spot can not be applied in this situation.

\section{Effect of scatterers}
\label{simp}

Since graphene is a two-dimensional material its transport properties are significantly affected by the presence of impurities.
We study the robustness of the Veselago lensing effect in the presence of scatterers.
Impurities are modeled by a change of on-site energy in our tight-binding calculations given by,
\begin{equation}
\label{egauss}
V_i =\sum_{n = 1}^N U_n \exp \left(-\frac{\left| r_i - r_n \right|^2}{2\xi _n^2}\right)
\end{equation}
where $N$ is the number of scattering sites.
The potential amplitude of the $n^\text{th}$ impurity, $U_n$, is randomly distributed within $\left[ -t, t \right]$, while the parameter $\xi_n$ is also randomly distributed within $\left[0, 5a_{cc} \right]$, where $t$ is a hopping parameter and $a_{cc}$ is the length of the carbon-carbon bond.
We use both $\xi_n < a_{cc}$ and $\xi_n > a_{cc}$ to include the effects of short- as well as long-range impurities.
The results of our calculations are shown in Figs. \ref{f7}(a) and (b) for armchair and zigzag system, respectively.
In both figures we alter the amount of impurities starting with $N = 150$ (red curve) up to $N = 1500$ (cyan curve) and compare the results with the clean system (blue curve).
Notice the destructive influence of the impurities on the peak at $E_F = 0.4$ eV.
Furthermore, unlike the clean case where we have a slow increase of transmission for $0< E_F <0.4$ eV (although with strong oscillations) this is not the case with the "dirty" systems.
In the latter case transmission decreases with the increase of the Fermi energy up to $E_F = V$.
However, it is interesting that at low energies certain "dirty" samples show higher transmission probability than the clean one.

\section{Conclusion}
\label{sconc}

In this paper we investigated the feasibility of Veselago lensing in graphene.
The effect was studied for two systems - a four terminal system with wide side-leads and a two terminal system with narrow leads.
This allowed us to distinguish the influence of edge and p-n interface scattering on the Veselago lensing effect.

In the case of the four-terminal structure we saw that the transmission has a distinctive peak at $E_F = V/2$ which is a consequence of electron focusing in graphene.
The current density plots confirmed that the injector is mirrored at the opposite side of the p-n junction resulting in a high transmission probability for both semiclassical and quantum calculations.
However, in the case of armchair edges at the p-n interface we found that the transmission peak is much lower.
We also investigated the influence of injector's width on the transmission maximum.
The calculations revealed that the highest transmission is achieved for narrow injectors and decreases continuously with the increase of the width.
Next, we examined the dependence of the transmission on the position of the p-n interface and found that the transmission approaches unity for the p-n interface placed at the center of the system.
The reason for this lies in the band gap that forms in the side-leads.
Furthermore, we found plateaus in the transmission that emerge due to the collimation of the beam after it has been focused in the focal spot.
Additionally, we considered a smooth p-n junction and observe changes in the transmission.
Simulations showed that all the main features were preserved, however smoothing of the potential broadened the focusing peak and decreases its height.

The study of the two-terminal system showed the influence of edge scattering on Veselago lensing.
For the case of a system with armchair edges we saw that the focusing peak decreased as compared with the case when the side-leads were added which is a result of the non-specular scattering on the edges.
In the case of the zigzag system the highest transmission probability was achieved for low Fermi energies when only one subband is occupied.
Our simulations showed that the width of the injector is not of importance in the case of semiclassical calculations, however this was not the case for the quantum mechanical tight-binding calculations.
The system with armchair edges showed that the transmission decreases with the increase of the injector's width.
We explained this with the fact that by increasing the width of the injector more electrons were scattered at the edges of the system which diminishes the electron focusing effect in graphene.
On the other hand, in the case of zigzag edges we saw that the transmission decreases in an oscillatory fashion with the width of the injector.

Finally, we studied the effect of impurities.
The calculations showed that the focusing peak around $E_F = V/2$ is very sensitive to the presence of impurities.
An increase of the contamination resulted in a decrease of the transmission for both armchair and zigzag systems.
However, we observed the exception at energies around the Dirac point where the transmission in certain "dirty" systems increased compared with the clean case.

\section{Acknowledgement}

This work was supported by the Flemish Science Foundation (FWO-Vl), the European Science Foundation (ESF) under the EUROCORES Program EuroGRAPHENE within the project CONGRAN and the Methusalem Foundation of the Flemish government.

\end{document}